\documentclass[aps,amsmath,twocolumn,amssymb,titlepage,10pt]{revtex4-1}
\usepackage[T1]{fontenc}
\usepackage[utf8]{inputenc}
\usepackage{amsmath}
\usepackage{braket}
\usepackage{amsfonts}
\usepackage{comment}
\usepackage{graphicx}
\usepackage{hyperref}
\usepackage[capitalize]{cleveref}
\usepackage{amssymb}
\usepackage{amsmath}
\usepackage{mathtools}
\usepackage{todonotes}

\begin{document}
\title{Anomalous scaling corrections and quantum phase diagram of the Heisenberg\\ antiferromagnet on the spatially anisotropic honeycomb lattice}
\author{Alexander Sushchyev}
\author{Stefan Wessel}
\affiliation{Institute for Theoretical Solid State Physics, RWTH Aachen University, JARA Fundamentals of Future Information Technology, and \\ JARA Center for Simulation and Data Science, 52056 Aachen, Germany}
\begin{abstract}
Using large-scale quantum Monte Carlo simulations, we determine the ground state phase diagram of the spin-1/2 antiferromagnetic Heisenberg model on the honeycomb lattice for the most generic case of three varying interaction strengths along the  different lattice directions. We identify continuous quantum phase transition lines that separate the long-range ordered antiferromagnetic regime from three distinct quantum-disordered phases, each  characterized by dominant dimer-singlet formations. The finite-size  behavior along these phase transition lines exhibits anomalously large corrections to scaling, which we relate to similar recent findings in certain dimerized quantum spin systems and to singular one-dimensional limits in the model parameter space. We also comment more generally on the non-universality of  critical cumulant ratios in anisotropic systems and  attempts to restore universality by varying the aspect ratio. 
\end{abstract}
\maketitle

\section{Introduction}\label{Sec:Introduction}

The honeycomb lattice has the lowest possible  coordination number ($z=3$)  among the eleven uniform Archimedean tilings (i.e., periodic tessellations of the plane by regular polygons such that all edge lengths are equal and every vertex looks alike)~\cite{Gruenbaum1977}. Nevertheless, its bipartite character  still allows for the stabilization of a N\'eel  antiferromagnetic (AFM)  ground state -- realized, e.g., in the spin-1/2 Heisenberg antiferromagnet with isotropic nearest-neighbor interactions, i.e, considering equal interaction strengths among all the bonds of the honeycomb lattice~\cite{Richter2004, Loew2009}. 
However, upon increasing the interaction strength along  one of the three inequivalent bond directions with respect to the other two, such a quantum spin system can be driven out of the AFM regime. 
More specifically, it has been shown that once the corresponding coupling ratio extends beyond about 1.735, the AFM  order is replaced by a non-magnetic, quantum disordered ground state, characterized by the dominant formation of singlets among the nearest-neighbor bonds with  the larger coupling strength~\cite{Jiang2009}. This way of dimerizing the honeycomb lattice is similar to models of dimerized quantum magnets on the square lattice. In the latter case, different dimerization patterns, such as columnar vs. staggered dimerization, have been intensively investigated recently~\cite{Matsumoto2001,Wang2006,Wenzel2008,Jiang2009,Jiang2012,Fritz2011,Yasuda2013,Ma2018}. In particular, for the specific case of a staggered dimerization large corrections to scaling have been reported at the quantum phase transition that separates the AFM phase from the quantum disordered regime~\cite{Wenzel2008,Fritz2011,Jiang2012,Yasuda2013,Ma2018}. These scaling corrections have been associated to   certain non-topological cubic terms in the effective low-energy field theory describing the quantum critical point~\cite{Fritz2011}. Apparently, they give rise to a  further weakly-irrelevant operator~\cite{Ma2018}, in addition to those expected for the three-dimensional classical Heisenberg universality class that generically characterizes the quantum phase transitions in such bipartite dimerized  spin systems~\cite{Chubukov1994}. 

\begin{figure}[t]
    \centering
    \includegraphics{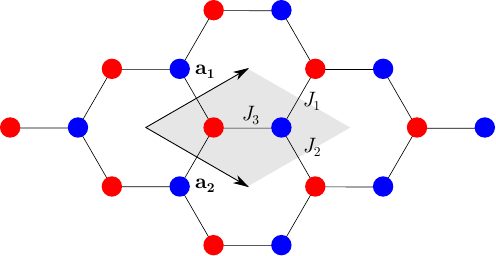}
    \caption{Illustration of the honeycomb lattice with its bipartite structure, as indicated by two  differently colored lattice sites. Also indicated is the unit cell spanned by the lattice vectors $\mathbf{a}_1$ and $\mathbf{a}_2$ as well as the exchange couplings $J_1$, $J_2$ and $J_3$ for the Heisenberg model along the three inequivalent bond directions. The shown finite lattice corresponds to $L=3$.}
    \label{Fig:lattice}
\end{figure}

Here, we report that similar anomalously large scaling correction also appear in the dimerized Heisenberg model on the honeycomb lattice. Moreover, in this case, we can link the estimated sub-leading scaling correction to the {\it leading} scaling correction in a special, one-dimensional limit  of the Heisenberg model on the honeycomb lattice. The latter is accessible  upon considering the most generic case, in which all three inequivalent nearest-neighbor bonds on the honeycomb lattice have a different coupling strength. In order to analyse this quantum spin system, we used large-scale quantum Monte Carlo (QMC) simulations, based on the stochastic series expansion (SSE) method~\cite{Sandvik1991,Sandvik1992,Sandvik1999}. 
From the QMC analysis of these quantum phase transitions, we then obtain the global ground state phase diagram of this most generic form of the nearest-neighbor spin-1/2 Heisenberg model on the honeycomb lattice with three different interaction strengths. Certainly, in view the abundant realizations of the honeycomb lattice geometry in magnetic compounds, this more generic quantum phase diagram  is of value by its own.

The remainder of this paper is organized as follows: In Sec.~\ref{Sec:Model}, we introduce the model Hamiltonian and comment on the employed QMC approach. Next, we present the ground state phase diagram as obtained from  QMC simulations in Sec.~\ref{Sec:Phases}. A detailed finite-size analysis of the anomalous scaling corrections is provided in
Sec.~\ref{Sec:Finite}, focusing on the Binder ratio~\cite{Binder1981,Ma2018}. We finally comment more generally on the non-universality of such critical cumulant ratios for anisotropic systems in Sec.~\ref{Sec:General}, before final conclusions are drawn in Sec.~\ref{Sec:Conclusions}.

\section{Model and method}\label{Sec:Model}

In the following, we consider the spin-1/2 Heisenberg model on the honeycomb lattice, taking into account three different exchange couplings  $J_1$, $J_2$ and $J_3$ along the three inequivalent nearest-neighbor bond directions, as indicated in Fig.~\ref{Fig:lattice} (we  consider  antiferromagnetic exchange interactions, $J_1,J_2,J_3\geq 0$). This system is described by the Hamiltonian
\begin{equation}
H=J_1 \sum_{\langle i,j \rangle_1}\mathbf{S}_i\cdot\mathbf{S}_j
+J_2 \sum_{\langle i,j \rangle_2}\mathbf{S}_i\cdot\mathbf{S}_j
+J_3 \sum_{\langle i,j \rangle_3}\mathbf{S}_i\cdot\mathbf{S}_j,
\end{equation}
where each of the three summations extends only over the parallel bonds corresponding to the labeling in Fig.~\ref{Fig:lattice}.
From previous studies it is well established that in the isotropic case, $J_1=J_2=J_3$, the ground state exhibits long-range AFM order, with a staggered magnetization $m_s^\mathrm{iso}$ that is reduced by about 54\% from its classical value in the perfect N\'eel state~\cite{Loew2009}. 
Upon enhancing one of the three exchange couplings, e.g., considering $J_1>J_2=J_3$, the AFM order gets suppressed, and for $J_1/J_2=J_1/J_3>1.735(1)$ the ground state becomes the aforementioned dimerized quantum disordered state  with a dominant singlet formation along the $J_1$ bonds~\cite{Jiang2009}. We denote this regime by D${}_1$ in the following. Correspondingly, upon increasing $J_2$ ($J_3$) with respect to $J_1=J_3$ ($J_1=J_2$), we obtain dimerized phases with dominant singlet formations along the $J_2$ ($J_3$) bonds, which we denote by D${}_2$ (D${}_3$). In the following, we  explore the phase diagram for the more general case in which the system is fully spatially  anisotropic, i.e., all three couplings may be varied independently. 

In order to extract the ground state properties of this system, we performed large-scale QMC simulations based on the SSE method with directed loop updates. We consider finite-size systems of rhombic shape as indicated in Fig.~\ref{Fig:lattice}, denoting by $L$ the extend of the  system in both the $\mathbf{a}_1$ and $\mathbf{a}_2$ directions, so that the number of spins $N=2L^2$. Furthermore, we employ periodic boundary conditions and  performed the QMC simulations at sufficiently low temperatures to probe ground state properties.
Based on the staggered structure factor
\begin{equation}
    S_{AF}=\frac{1}{N}\sum_{i,j} \epsilon_i \epsilon_j \mathbf{S}_i\cdot \mathbf{S}_j,
\end{equation}
where $\epsilon_i=\pm 1$, depending on the sublattice on  which $\mathbf{S}_i$ is located, we can estimate the staggered magnetization 
\begin{equation}
    m_s=\sqrt{S_{AF}/N}
\end{equation}
after an extrapolation to the thermodynamic limit. 
Furthermore, as detailed below, we can extract the phase transition lines out of the AFM regime upon measuring the Binder ratio~\cite{Binder1981,Jiang2009,Ma2018}
\begin{equation}
R=\frac{\langle M_s^4\rangle}{\langle M_s^2\rangle^2},
\end{equation}
where $M_s=\sum_i \epsilon_i S^z_i$ denotes the staggered moment.
In the following section, we first present the resulting phase diagram of the Hamiltonian $H$, before we discuss the QMC-based determination of the phase boundaries  in Sec.~\ref{Sec:Finite}.

\begin{figure}[t]
    \centering
    \includegraphics{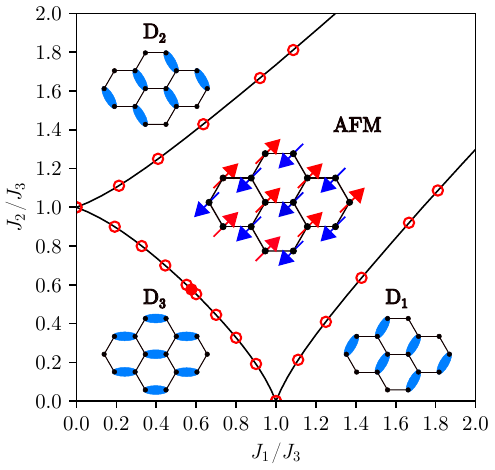}
    \caption{Ground state phase diagram of the spin-1/2 Heisenberg model on the honeycomb lattice, expressed in the two parameter ratios $J_1/J_3$ and $J_2/J_3$. 
   Circles indicate results from QMC simulations, while the solid line is a guide to the eye.  Illustrated within each  quantum disordered regime is the  dominant dimerization pattern.}
    \label{Fig:Phasediag1}
\end{figure}

\section{Ground state phase diagram}\label{Sec:Phases}

\begin{figure}[t]
    \centering
    \includegraphics{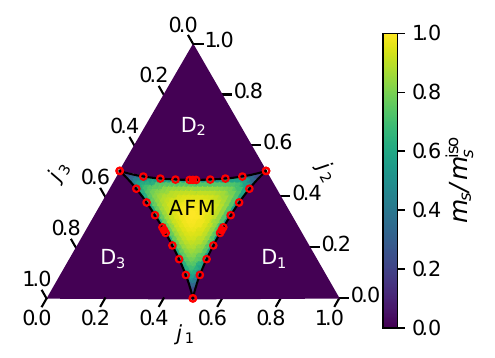}
    \caption{Ground state phase diagram of the spin-1/2 Heisenberg model on the honeycomb lattice, expressed in barycentric coordinates. Circles indicate results from QMC simulations, while the solid line is a guide to the eye.  Also shown inside the AFM regime is the magnitude of the staggered magnetization $m_s$, normalized to its maximum value $m_s^\mathrm{iso}$ in the isotropic case (located at the center of the triangle). }
    \label{Fig:Phasediag2}
\end{figure}

We first consider a parametrization in units of the coupling $J_3$, i.e., tuning the ratios $J_1/J_3$ and $J_2/J_3$. From the finite-size analysis of the QMC data,  detailed in the next section, we then obtain the ground state phase diagram 
shown in Fig.~\ref{Fig:Phasediag1}, which also illustrates the different ground states. An alternative form of the phase diagram, adapted  to its symmetries, is obtained upon expressing it in terms of barycentric coordinates, defined as $j_i=J_i/(J_1+J_2+J_3)$, such that $j_1+j_2+j_3=1$. The resulting phase diagram is shown in Fig.~\ref{Fig:Phasediag2}. Included in this figure is the value of the staggered magnetization $m_s$ inside the AFM phase, normalized to its maximum value $m_s^\mathrm{iso}$ in the isotropic case (located at the center of the triangle). 

From these ground state phase diagrams, we find that the three distinct quantum disordered regimes  D${}_1$,  D${}_2$, and  D${}_3$ are generically separated by the AFM phase, as they meet only pairwise at singular points that are located on  special lines where one coupling strength vanishes. For example, for $J_1=0$, the system decouples into zig-zag chains (cf. Fig~\ref{Fig:lattice}) that are aligned along the $\mathbf{a}_2$ direction. If $J_2=J_3$, these spin-1/2 Heisenberg chains have a  gapless,  quantum critical ground state, and an arbitrarily weak interchain coupling $J_1>0$
drives the system into the AFM regime~\cite{Giamarchi2004}. However, for $J_2>J_3$ ($J_3>J_2$) the isolated chains have a gapped ground state, with dominant singlet formations along the $J_2$ ($J_3$) bonds, corresponding to the one-dimensional limit of the D${}_2$ (D${}_3$) phase~\cite{Giamarchi2004}. In the latter case, a finite interchain coupling $J_1>J_1^c>0$ is therefore required to close the excitation gap and drive the system into the AFM regime. This quasi-one-dimensional physics, along with the corresponding cases for $J_2=0$ and $J_3=0$, gives rise to the curved triangular shape of the AFM regime in Fig.~\ref{Fig:Phasediag2}.

In the following section, we  provide a detailed finite-size analysis of the Binder ratio, based on which the above phase boundaries were obtained. In particular, we show that the Binder ratio exhibits  anomalous finite-size corrections which can be linked to the above mentioned singular points. 

\section{Finite-size analysis}\label{Sec:Finite}

In order to determine the stability range of the AFM regime, we typically fixed in the QMC simulations the value of $J_2/J_3$, and varied $J_1/J_3$, i.e., we probed the system along horizontal cuts in Fig.~\ref{Fig:Phasediag1}. Indeed, we can furthermore restrict ourselves to the regime $J_1 \leq J_2 \leq J_3 $, as  other sections of the AFM boundary can be obtained after appropriate relabelings of  the coupling constants.  

\begin{figure}[t]
    \centering
    \includegraphics{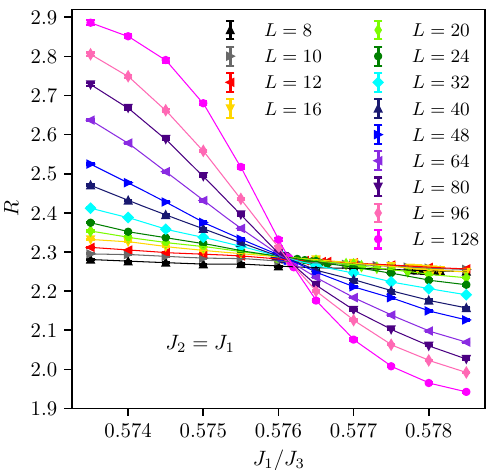}
    \caption{Binder ratio $R$ as a function of $J_1=J_2$ for different system sizes $L$. }
    \label{Fig:Riso}
\end{figure}

However, here we first consider the symmetric line along which $J_1=J_2$. 
Figure~\ref{Fig:Riso} shows the dependence of the Binder ratio $R$ on the coupling ratio $J_1/J_3$ along this diagonal. As a dimensionless quantity, $R$ is particularly  useful for finite-size studies of (quantum) critical phenomena, since to leading order it 
exhibits a crossing point at the (quantum) critical point when considering different finite-size systems with a fixed  spatial (as well as -- for quantum phase transitions -- the space-time) aspect ratio. Here, we consider $L\times L$ lattices and fix the inverse temperature $\beta=2L$ (in units of $1/J_3$)~\cite{Ma2018}. 

\begin{figure}[t]
    \centering
    \includegraphics{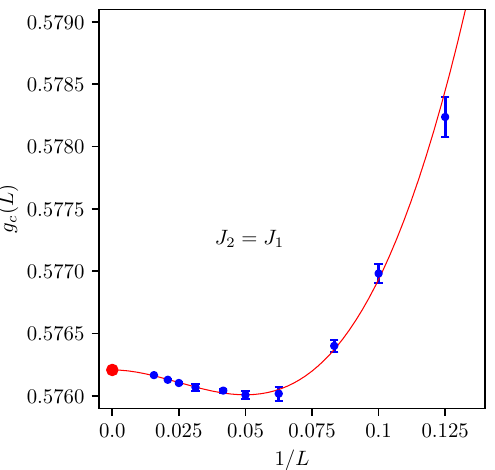}
    \caption{Crossing point values of  $g=J_1/J_3$ for the Binder ratio $R$ for $J_2=J_1$ between system sizes $L$ and $2L$. Also included is a fit to the finite-size scaling form (\ref{Eq:Scalingg}).}
    \label{Fig:J1isocross}
\end{figure}

\begin{figure}[t]
    \centering
    \includegraphics{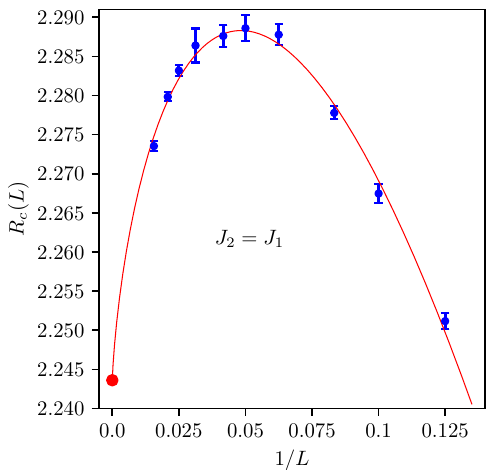}
    \caption{Crossing point values of the Binder ratio $R$ for $J_2=J_1$ between system sizes $L$ and $2L$. Also included is a fit to the finite-size scaling form (\ref{Eq:ScalingR}).}
    \label{Fig:Risocross}
\end{figure}

The QMC data for $R$ in Fig.~\ref{Fig:Riso} indeed cross for different system sizes near $J_1/J_3\approx 0.576$. However, closer inspection reveals that there is a systematic drift of the crossing points between different system sizes, and we therefore  need to examine in more detail the crossing point values of $g=J_1/J_3$ and $R$ between system sizes $L$ and $2L$ in Fig.~\ref{Fig:J1isocross} and Fig.~\ref{Fig:Risocross}, respectively. 

A striking feature of the finite-size dependence of both quantities is a strongly nonmonotonic behavior,  similar to the one observed for the staggered dimer square-lattice model (but not for the columnar dimer case) in Ref.~\cite{Ma2018}. This similarity may not come as a surprise if one compares the two lattice structures: Also for the honeycomb lattice, the stronger bonds (in this case the $J_3$ bonds) are arranged in a staggered fashion (but with fewer interdimer couplings than for the staggered dimer square lattice). 
We note that such a nonmonotonic behavior was however not reported in Ref.~\cite{Jiang2009}, possibly due to (i) a lower resolution in scanning the coupling ratio near the quantum critical point for the large system sizes than here, and/or (ii) the different finite-size lattice geometry employed in that study.

In Ref.~\cite{Ma2018}, a careful finite-size analysis of such nonmonotonic scaling corrections was performed. It was concluded that they are best accounted for by considering {\it two} subleading correction terms to the leading scaling form,
\begin{eqnarray}
g_c(L)&=&g_c+L^{-1/\nu} (a_1 L^{-\omega_1}+a_2 L^{-\omega_2}),\label{Eq:Scalingg}\\ 
R_c(L)&=&R_c+b_1 L^{-\omega_1}+b_2 L^{-\omega_2},\label{Eq:ScalingR}
\end{eqnarray}
where $g$ denotes the tuning parameter that drives the quantum phase transition (here, $g=J_1/J_3$), $\nu$ is the correlation length scaling exponent, and $\omega_1$ and $\omega_2$ two correction exponents (the $a_i$, $b_i$ are nonuniversal prefactors). For the quantum phase transition under consideration here as well as in Ref.~\cite{Ma2018}, $1/\nu=1.406$ and $\omega_1=0.78$ can be taken from the classical values of the three-dimensional Heisenberg universality 
class~\cite{Guida1998,Hasenbusch2001,Campostrini2002} (note that besides $\nu$, also the correction exponent $\omega_1$ is universal within a given universality class). Furthermore,  Ref.~\cite{Ma2018} estimated a value of $\omega_2\approx 1.25$ for the staggered dimer square-lattice model. This value is greater than $\omega_1$ and smaller than $2\omega_1$, i.e., $\omega_1$ is still the leading scaling correction, related to the three-dimensional classical Heisenberg universality class. The further correction term, related to $\omega_2$, was argued to steam from the cubic term in the effective action of the field theory describing the quantum critical point~\cite{Fritz2011,Ma2018}.

\begin{figure}[t]
    \centering
    \includegraphics{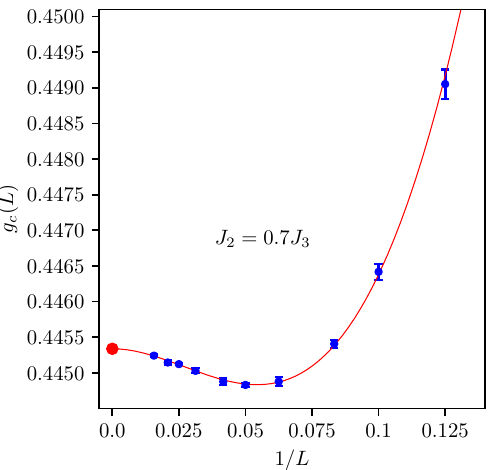}
    \caption{Crossing point values of  $g=J_1/J_3$ for the Binder ratio $R$ for $J_2=0.7 J_3$ between system sizes $L$ and $2L$. Also included is a fit to the finite-size scaling form (\ref{Eq:Scalingg}).}
    \label{Fig:J1cross}
\end{figure}

\begin{figure}[t]
    \centering
    \includegraphics{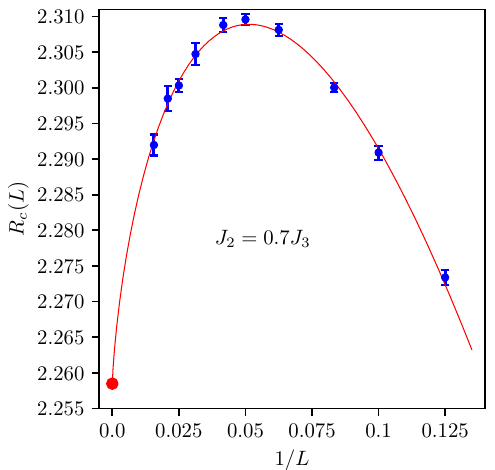}
    \caption{Crossing point values of the Binder ratio $R$ for $J_2=0.7 J_3$ between system sizes $L$ and $2L$. Also included is a fit to the finite-size scaling form (\ref{Eq:ScalingR}).}
    \label{Fig:Rcross}
\end{figure}

We find that the above extended scaling ansatz also fits well (i.e., with values of $\chi^2/\mathrm{d.o.f.}\approx 1$) to the anomalous, nonmonotonic behavior that we observe for the honeycomb lattice model: Fits to the QMC data based on this ansatz are included in  Fig.~\ref{Fig:J1isocross} and Fig.~\ref{Fig:Risocross} (unfortunately, the accessible accuracy on the estimated crossing points does not allow us to robustly estimate the exponents independently).
Based on these fits, we furthermore obtain an estimate for the critical coupling $J_1^c=0.57620(1) J_3$ (shown in Fig.~\ref{Fig:Phasediag1} and Fig.~\ref{Fig:Phasediag2} by the full circle) and $R_c=2.244(1)$, which are in accord with earlier estimates~\cite{Jiang2009} along the symmetric line (the stated uncertainties refer to the performed fits and do not account for the unknown uncertainty in the value of $\omega_2$ estimated in Ref.~\cite{Ma2018}). 
We find that the above anomalous finite-size  behavior persists also away from the symmetric line. As a further example, we discuss here $J_2=0.7J_3$. 
In this case, fits to the data shown in Fig.~\ref{Fig:J1cross} and ~\ref{Fig:Rcross} lead to $J_1^c=0.44533(1)J_3$ and $R_c=2.258(1)$, respectively.


\begin{figure}[t]
    \centering
    \includegraphics{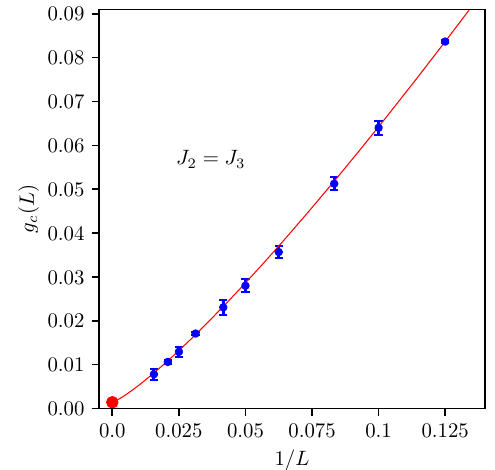}
    \caption{Crossing point values of  $g=J_1/J_3$ for the Binder ratio $R$ for $J_2=J_3$ between system sizes $L$ and $2L$. Also included is a fit to the finite-size scaling form (\ref{Eq:Scalingg1}).}
    \label{Fig:J1cross1}
\end{figure}

\begin{figure}[t]
    \centering
    \includegraphics{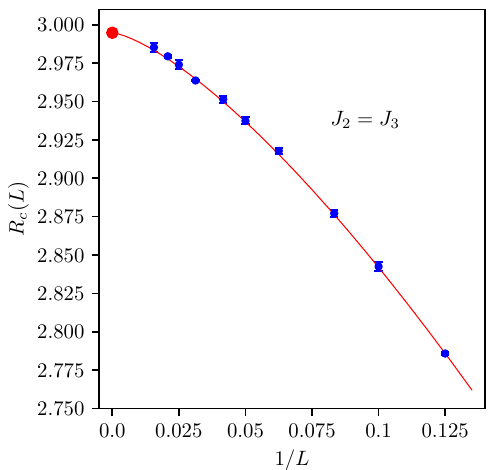}
    \caption{Crossing point values of the Binder ratio $R$ for $J_2=J_3$ between system sizes $L$ and $2L$. Also included is a fit to the finite-size scaling form (\ref{Eq:ScalingR1}).}
    \label{Fig:Rcross1}
\end{figure}

By contrast, we observe a rather distinct finite-size scaling behavior if we fix the coupling ratio to $J_2=J_3$. In this case, the data shown in  Fig.~\ref{Fig:J1cross1} and ~\ref{Fig:Rcross1} does not feature the above nonmonotonic behavior, instead  it fits well to scaling forms with only a single   exponent each, which we denote by $p$ and $\omega$, respectively,
\begin{eqnarray}
g_c(L)&=&g_c+a\: L^{-p} ,\label{Eq:Scalingg1}\\ 
R_c(L)&=&R_c+b\: L^{-\omega}.\label{Eq:ScalingR1}
\end{eqnarray}
The extracted value of $J_1^c=0.001(1)J_3$ in this case is consistent with zero and thus in accord with the quasi-one-dimensional physics mentioned in the previous section. Indeed, also the extracted value of $R_c=2.996(2)$ is compatible with the value of $R_c=3.05(5)$ that we obtain from QMC simulations performed for an isolated spin-1/2 Heisenberg chain (cf. App.~\ref{Sec:app}). 

Furthermore, we find that the estimated values of the correction exponent $\omega\approx 1.40(4)$ as well as $p \approx 1.21(5)$ are comparable to the above cited estimate for $\omega_2$, suggesting that  the subleading scaling corrections, captured by the $L^{-\omega_2}$ term, may arise due to a simple crossover effect: For small system sizes, the finite-size behavior is still affected by the influence of the specific fixed-point (in a renormalization-group sense), corresponding to the decoupled spin-1/2 Heisenberg chains. Only on sufficiently large system sizes does the actual asymptotic scaling of the three-dimensional classical Heisenberg universality class prevail. 
This scenario provides a natural explanation for the nonmonotonic finite-size behavior observed above (a crossover scale of $L_c\approx 20$ can then be estimated based on the local extrema in the finite-size crossing points). We comment further on this point and its relation to previous work in the final Sec.~\ref{Sec:Conclusions}. 

\begin{figure}[t]
    \centering
    \includegraphics{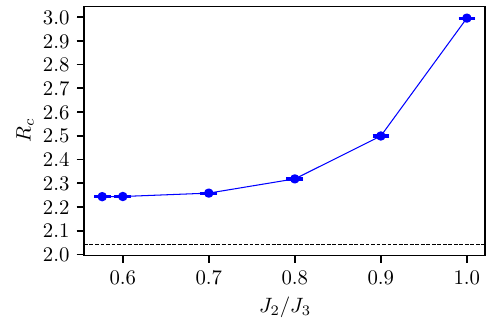}
    \caption{Critical Binder ratio $R$ as a function of $J_2$ along the considered phase transition line. The dashed line indicates the value of the critical Binder ratio $R_c^\mathrm{iso}=2.0437...$ of the isotropic three-dimensional classical Heisenberg model.}
    \label{Fig:Rall}
\end{figure}

In Fig.~\ref{Fig:Rall}, we summarize the extrapolated values of the critical Binder ratio $R_c$ as a function of $J_2$ along the considered branch of the AFM phase boundary line. We observe a systematic dependence of the value of $R_c$ on the coupling ratios, along with the approach towards $3$ for $J_2=J_3$. In particular, the values that we obtain for $R_c$ are all substantially larger than the critical value $R_c^\mathrm{iso}=2.0437...$ of the isotropic  three-dimensional classical Heisenberg model.  This property reflects  the fact that the correlations in the considered quantum spin systems are not isotropic, a property that indeed affects the value of the critical Binder ratio~\cite{Chen2004,Selke2005}. We elaborate further on this point in the following section. 

\section{General remark on anisotropic systems}\label{Sec:General}

The fact that certain dimensionless quantities such as the critical Binder ratio $R_c$ are not universal, but depend on the anisotropy of the correlations~\cite{Chen2004,Selke2005} is indeed  well established by now. It has been suggested~\cite{Yasuda2013} that the isotropic value $R_c^\mathrm{iso}$ may still be restored upon considering finite systems with an appropriate aspect ratio (even though such a restoration has been argued to not be crucial for the analysis of, e.g., the universal properties of the critical point~\cite{Ma2018}). More specifically, for the simple case of an underlying square lattice geometry, one would consider a finite rectangle with $L_x$ ($L_y$) unit cells in the $x$ ($y$) direction, such that by tuning the aspect ratio $\rho=L_y/L_x$, the critical Binder  ratio $R_c(\rho)$ takes on the isotropic value $R_c(\rho^*)=R_c^\mathrm{iso}$ for an appropriate value of $\rho=\rho^*$, even when $R_c(1)\neq R_c^\mathrm{iso}$ (for a quantum critical system, the inverse temperature has to be adapted correspondingly as well~\cite{Yasuda2013}). Such restorations have indeed been demonstrated for various classical and quantum models for which the principal axes of the correlation function align with the axes of the underlying square lattice~\cite{Kamieniarz1993,Selke2005,Selke2007,Yasuda2013}.  
For a general anisotropic two-dimensional system, the angular dependence of the correlations is given by the angle $\Omega$, specifying  the orientation of the two principal axes, and the ratio $q$ of the two principal correlation lengths upon approaching criticality~\cite{Dohm2019}. 
The above special cases correspond to $\Omega=0$ or $\pi/2$. For example, in the first case, the value of $\rho^*$ is fixed by the condition that $q\rho^*=1$ (e.g., for $q=2$, such that the correlation length in the $x$ direction is twice the one along the $y$ direction, this yields $\rho=1/2$, i.e., $L_x=2 L_y$). Such systems were termed "virtual isotropic"~\cite{Matsumoto2001,Yasuda2013}.

It is not clear, (i) whether such a restoration is also feasible at all in the more generic case, where $\Omega$ takes on different values, i.e., when the principle axes are not aligned with the lattice directions, and, if so,  (ii) how to physically interpret the  value of $\rho^*$ in that case. 
In the following, we demonstrate explicitly for a basic classical model  that such a  restoration of the isotropic value is apparently still possible -  however, the value of $\rho^*$ appears in this case to be unrelated to the orientation of the underlying correlations. 

\begin{figure}[t]
    \centering
    \includegraphics{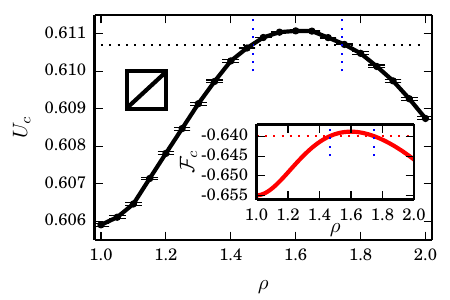}
    \caption{Dependence of the critical Binder cumulant $U_c$ (main panel) and the critical amplitude of the singular free energy density  $\mathcal{F}_c$ (inset)  on the aspect ratio $\rho$ for the anisotropic Ising model with $E_3=3E$. Isotropic values are indicated by horizontal lines. Vertical lines indicate the two values of $\rho^*$. The left inset illustrates the unit cell with $E_1$, $E_2$, $E_3$ along horizontal, vertical and diagonal directions, respectively.}
    \label{Fig:U}
\end{figure}

For this purpose, we consider the triangular-lattice Ising model~\cite{Stephenson2004, Dohm2019, Dohm2021, Dohm2021a}, defined by
\begin{equation}
\label{IsingH}
H^{{\text{\rm{Is}}}}\! =\!-\!\!\sum_{i}\big [E_1\sigma_{i} \sigma_{i+\hat{x}}+E_2\sigma_{i} \sigma_{i+\hat{y}}  +E_3\sigma_{i} \sigma_{i+\hat{x}+\hat{y}}\big],
\end{equation}
where $\sigma_{i}=\pm1$ reside on a square lattice with horizontal, vertical, and (up-right) diagonal couplings $E_1, E_2$, $E_3\geq 0$ (the unit cell is illustrated in the inset of Fig.~\ref{Fig:U}). The condition for the critical temperature $T_c$, separating the low-$T$ ferromagnetic phase from the paramagnetic regime, reads 
$\hat S_1 \hat S_2+\hat S_2 \hat S_3+\hat S_3 \hat S_1=1$, where
$\hat S_\alpha= \sinh (2 E_\alpha/T_c)$~\cite{Houtappel1950}.
In the following, we focus on the symmetric case $E_1=E_2=E$, for which $\Omega=\pi/4$ and $q=1/\hat S_3$~\cite{Dohm2019,Dohm2021a}. Furthermore, the correlations decay equally along the $x$ and $y$ directions in this case.

In Fig.~\ref{Fig:U} we report the $\rho$-dependence of the critical Binder cumulant~\cite{Binder1981} 
\begin{equation}
    U_c=1-\frac{1}{3}R_c, \quad R_c=\frac{\langle M^4\rangle}{\langle M^2\rangle^2},
\end{equation}
where $M=\sum_i \sigma_i$,
for the specific case of $E_3=3E$, as obtained using Wolff-cluster Monte Carlo simulations~\cite{Wolff1989} at $T_c$, extracted for the thermodynamic limit. The value  of $U_c$ obtained for $\rho=1$ agrees with an earlier study~\cite{Selke2005}. In contrast to the standard Ising model ($E_1=E_2$, $E_3=0$), for which $U_c(\rho)$ monotonously decreases for $\rho>1$ from the isotropic value $U_c^\mathrm{iso}=0.61069...$ at $\rho=1$~\cite{Kamieniarz1993}, here it exhibits a nonmonotonous behavior. In particular, the isotropic value is recovered at two specific values,  $\rho^*\approx 1.47$ and $1.74$. While there is no apparent relation of these numbers to the value of $q=2.82843...$, we find that within the numerical precision they do agree with the values of $\rho$ for which the critical amplitude of the singular free energy density $\mathcal{F}_c(\rho)$ recovers its isotropic value $\mathcal{F}_c^{\mathrm{iso}}=-0.639912...$~\cite{Ferdinand1969}, cf. the inset of Fig.~\ref{Fig:U}. For the latter quantity we  can extract $\rho^*$ using exact conformal field theory-based expressions for $\mathcal{F}_c$~\cite{Dohm2021,Dohm2021a}. For $E_3=3E$,  we obtain $\rho^*=1.4683...$ and $1.7433...$, respectively. 

We made the above observations also at other ratios $E_3/E$ (see also Ref.~\cite{Selke2007} for the case $E_3=E$), yielding different values of $\rho^*$, and  suggesting that interestingly, (i) it is feasible within the considered anisotropic Ising model to restore the isotropic values of both the critical Binder cumulant and the critical amplitude of the singular free energy density upon varying the aspect ratio $\rho$,  (ii) the corresponding value(s) of $\rho^*$ for both quantities are very close  -- if not equal --, and (iii) the value(s) of $\rho^*$ are not obviously related to geometric properties of the anisotropic correlation functions for $\Omega\neq 0, \pi/2$. It would certainly be interesting to examine these points in more detail in the future, both for the triangular-lattice Ising model and on more general grounds. 

For the purpose of the present discussion, the relevant conclusion is: Once the principal axes are not aligned to the underlying lattice directions, one cannot expect to be able to relate, as in previous cases~\cite{Kamieniarz1993,Selke2005,Yasuda2013}, the value of $\rho^*$ directly to the correlation length ratio along the lattice directions (which equals 1 in the above considered Ising model cases, since $E_1=E_2$), nor to the ratio $q$ of the two principal correlation lengths. In particular, it is thus not clear, how the procedure proposed in Ref.~\cite{Yasuda2013} for the automated  location of $\rho^*$ can be generalized to the fully anisotropic case.  

\section{Conclusions}\label{Sec:Conclusions}
Based on large-scale quantum Monte Carlo simulations we determined the ground state phase diagram of the spin-1/2 Heisenberg model on the honeycomb lattice with spatially fully anisotropic nearest-neighbor exchange interactions. In addition to the AFM phase, three quantum disordered regimes have been identified, each characterized by a dominant dimer singlet formation along the strongest bonds. 
The disordered phases are generically separated by the AFM regime, but meet pairwise at three singular points in the extended parameter domain. 

The generic quantum phase transition along the boundary line of the AFM regime exhibits  nonmonotonic anomalous finite-size corrections, similarly to the previously investigated staggered dimer model~\cite{Ma2018}. Based on a scaling ansatz with two irrelevant scaling correction terms, this behavior could be well captured.  The additional scaling correction  is subdominant to the leading finite-size correction that relates to the three-dimensional O(3) universality class. 
Upon approaching the aforementioned singular points, the system decouples into parallel arrays of spin-1/2 Heisenberg (zig-zag) chains, for which a similar finite-size correction term was identified to {\it leading} order. This suggests a crossover scenario to explain the nonmonotonic finite-size effects.

It may be  worthwhile to mention that the cubic interaction term that was previously proposed to underlie  the anomalous scaling observed in the staggered dimer model~\cite{Fritz2011} has a  formal similarity  with the topological $\theta$-term that is characteristic for the gapless state of the spin-1/2 Heisenberg chain~\cite{Giamarchi2004}. For the future it would thus be interesting to explore, if a similar link between the subleading scaling correction and the behavior at the one-dimensional singular points that we observe here also emerges in appropriate generalizations of the staggered dimer model. 

Finally, we pointed out that previously proposed schemes~\cite{Yasuda2013} for the automated restoration of isotropic values of cumulant ratios upon varying the aspect ratio are not directly applicable to generic anisotropic systems, even in cases for which such a restoration may be feasible. Certainly, more work is required to explore the consequences of this observation. 

\section*{Acknowledgements}
We thank Nils Caci, Florian Kischel, Francesco Parisen Toldin, and Adrien Reingruber for useful discussions. Furthermore, 
we acknowledge support by the Deutsche Forschungsgemeinschaft (DFG) through Grant No.~WE/3649/4-2 of the FOR 1807 and through RTG 1995, and thank the IT Center at RWTH Aachen University for access to computing time through the JARA Center for Simulation and Data Science.

\appendix
\section{Binder ratio of the spin-1/2 Heisenberg chain}\label{Sec:app}
\begin{figure}[t]
    \centering
\includegraphics[width=0.5\textwidth]{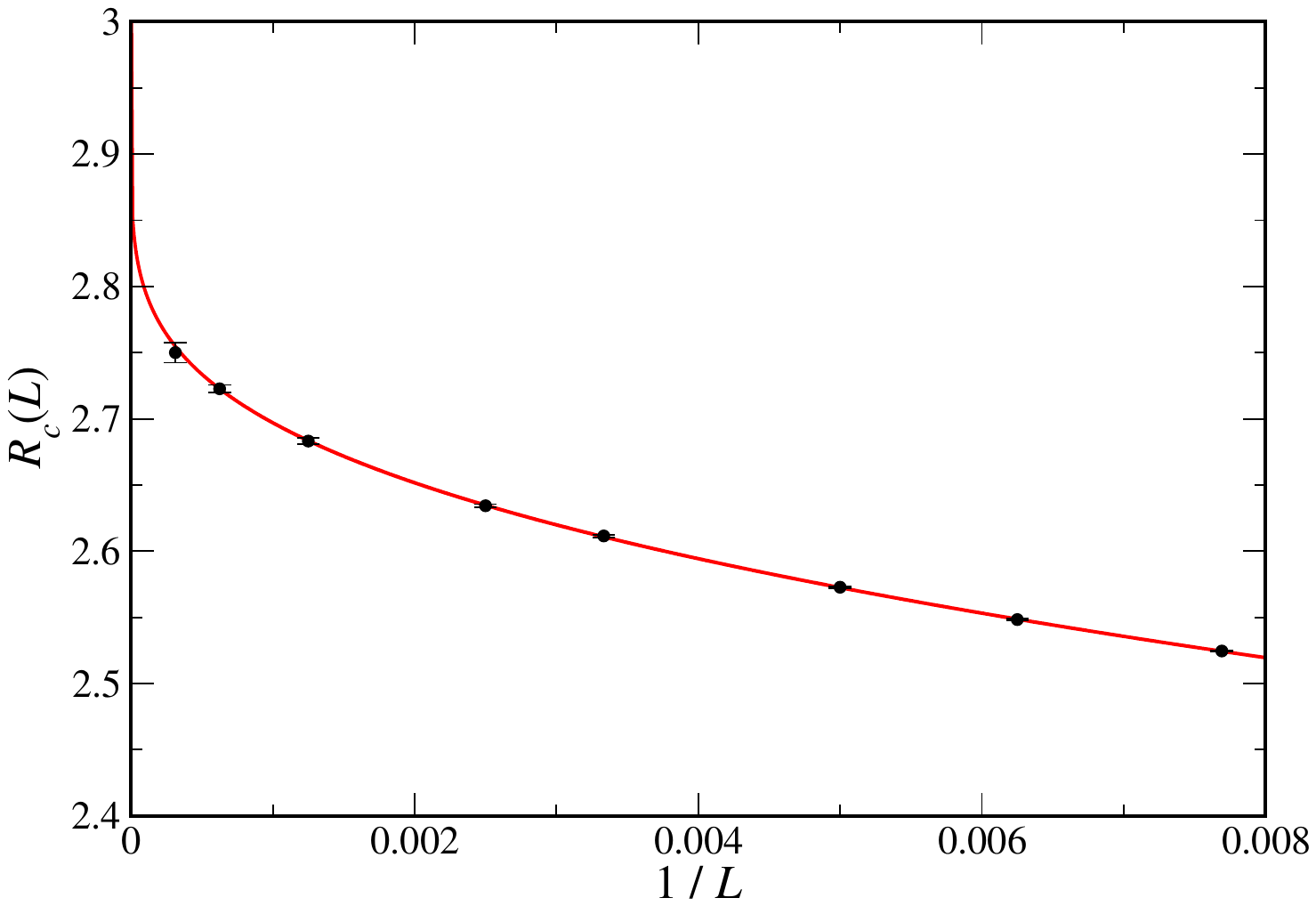}
    \caption{Critical Binder ratio $R_c(L)$ for the spin-1/2 Heisenberg chain along with a fit to the finite-size scaling form (\ref{Eq:Scaling1D}).}
    \label{Fig:1D}
\end{figure}
We  performed QMC simulations for an isolated spin-1/2 Heisenberg chain in order to directly estimate the value of $R_c$ at the singular points of the phase diagram. The results of these simulations are shown in Fig.~\ref{Fig:1D}, including data $R_c(L)$ for 
chains with $L$ spins and periodic boundary conditions up to $L=3200$, scaling the inverse temperature $\beta J=L$ in units of the nearest-neighbor exchange coupling $J$ along the chain. 
We obtain a monotonous increase of $R_c(L)$ with increasing system size that fits well to a logarithmic finite-size dependence,
\begin{equation}\label{Eq:Scaling1D}
 R_c - R_c(L) \propto [\ln(L)]^{-q},
\end{equation}
and from performing a fit to the QMC data, we obtain the estimates $R_c=3.05(5)$ and $q=1.1(1)$, respectively. Such a logarithmic finite-size behavior of $R_c(L)$ can be expected given that the asymptotic spin-spin correlation function of the spin-1/2 Heisenberg chain is well known to decay with distance $r$ as $\sqrt{\ln(r)}/r$ ~\cite{Giamarchi2004}, i.e., including a logarithm as well. We are  not aware of any previous analytic prediction for the finite-size scaling form of the Binder ratio in this system. 

\bibliography{references.bib}
\end{document}